\documentstyle[clbiba]{article} 
\input{psfig}

\newenvironment{outline}{\begin{list}{\arabic{enumi}.}{\usecounter{enumi}\setlength{\itemsep}{0ex}\setlength{\parsep}{0ex}}}{\end{list}}

\begin{document}
\bibliographystyle{abbrv}
\thispagestyle{empty}

\hrule

\vspace{1.5ex}\noindent
{\LARGE\bf Developing numerical libraries in Java}

\vspace{3.5ex}\noindent
RONALD F.~BOISVERT$^1$, JACK J.~DONGARRA$^2$, ROLDAN POZO$^1$, \\
KARIN A.~REMINGTON$^1$ AND G.W.~STEWART$^{1,3}$ 

\vspace{2ex}\noindent
$^1${\it Mathematical and Computational Sciences Division,
     Information Technology Laboratory, 
     National Institute of Standards and Technology, 
     Gaithersburg, MD 20899 USA \\
     (email: \{boisvert,pozo,karin\}@nist.gov)}
  
\vspace{2ex}\noindent
$^2${\it Computer Science Department,
     University of Tennessee at Knoxville, 
     Knoxville, TN 37996, and
     Oak Ridge National Laboratory, Oak Ridge, TN \\
     (email: dongarra@cs.utk.edu)}

\vspace{2ex}\noindent
$^3${\it Department of Computer Science,
     University of Maryland,
     College Park, MD 20742 USA \\
     (email: stewart@cs.umd.edu)}

\vspace{3ex}
\hrule

\vspace{5ex}\noindent
{\bf SUMMARY \vspace{0.5ex} \\
The rapid and widespread adoption of Java has created a demand for reliable and
reusable mathematical software components to support the growing number of
compute-intensive applications now under development, particularly in science
and engineering.  In this paper we address practical issues of the Java
language and environment which have an effect on numerical library design and
development.  Benchmarks which illustrate the current levels of performance of
key numerical kernels on a variety of Java platforms are presented.  Finally, a
strategy for the development of a fundamental numerical toolkit for Java is
proposed and its current status is described.
}


\section{INTRODUCTION}

Mathematical software libraries were introduced in the 1960s both to promote
software reuse and as a means of transferring numerical analysis and
algorithmic expertise to practitioners.  Many successful libraries have since
been developed, resulting in a variety of commercial products, as well as
public repositories of reusable components such as {\em netlib} and the Guide
to Available Mathematical Software \cite{Boisvert:1996:DSD}.

Library users want components which run fast, are easily moved among computing
platforms, invariably produce the right answer, and are easy to
understand and integrate with their applications.  Thus, efficiency,
portability, reliability and usability are of primary concern to library
developers.  Unfortunately, these properties are often competing, portability
and reliability often taking a toll on performance, for example.  Hence, the
development of high quality portable mathematical software libraries for widely
differing computing environments continues to be a challenging task.

Java technology \cite{Gosling:1996:JLS,Lindholm:1996:JVM} is leading to a
revolution in network-based computing. One of the main reasons for this is
the promise of new levels of portability across a very wide range of
platforms. Java is only beginning to affect the scientific computing
world. Some of the barriers to its adoption in this domain are the
perception of inadequate efficiency, language constraints which make
mathematical processing awkward, and lack of a substantial existing base
of high quality numerical software components. 

In this paper we assess the suitability of the Java language for the
development of mathematical software libraries. We focus on features of
the language and environment which may lead to awkward or inefficient
numerical applications. We present case studies illustrating the
performance of Java on key numerical kernels in a variety of environments.
Finally, we outline the Java Numerical Toolkit\footnote{\tt
http://math.nist.gov/jnt/} (JNT), which is meant to provide a base of
computational kernels to aid the development of numerical applications and
to serve as a basis for reference implementations of community defined
frameworks for computational science in Java. 


\section{NUMERICAL COMPUTING IN JAVA}

Java is both a computer language and a run-time environment. The Java
language \cite{Gosling:1996:JLS} is an object-oriented language similar
to, but much simpler than, C++. Compilers translate Java programs into
bytecodes for execution the Java Virtual Machine (JVM)
\cite{Lindholm:1996:JVM}. The JVM presents a fixed computational
environment which can be provided on any computer platform. 

The resulting computing environment has many desirable features: a simple
object-oriented language, a high degree of portability, a run-time system that
enforces array bounds checking, built-in exception handling, and an automated
memory manager (supported by a garbage collector), all of which lead to more
reliable software.

In this section we review key features of the Java language, assessing their
effect on both performance and convenience for use in numerical computing.  In
doing this we point out a number of deficiencies in the language.  It is
important to note, however, that many of Java's desireable features, such as
its portability, are derived from the JVM rather than the language itself.
Other languages can be compiled into Java bytecodes for execution by the JVM,
and several compilers for Java extensions are under development.\footnote{Many
of these are listed at \verb+http://grunge.cs.tu-berlin.de/$\sim$tolk/vmlanguages.html+.} Precompilers
which translate other languages, including C++, into pure Java are also under development.  If such tools achieve a high enough level of maturity and support,
they too can provide the basis for a Java-based development environments for
scientific computing.

\subsection{Arithmetic}


The idea that results produced on every JVM should be bitwise identical
\cite{Gosling:1996:JLS} on all platforms threatens the usability of Java for
high performance scientific computing.  While there may be some scientific
applications where such certainty would be useful, its strict implementation
could severely degrade performance on many platforms.  Systems with hardware
that support extended precision accumulators (which enhance accuracy) would be
penalized, for example, and certain code optimizations (including run-time
traslation to native code) would be disallowed.

It is also unfortunate that Java has not provided programmers with full access
to the facilities of IEEE floating-point arithmetic.  Programmers do not have
control of the rounding mode, for example (although this is rarely available in
high level languages).  Also, the ability to (optionally) have floating-point
arithmetic throw exceptions (on the generation of a NaN, for example), would
simplify coding and debugging.

\subsection{Complex arithmetic}

Complex arithmetic is essential in scientific computing. Java does not
have a complex data type, although this is not a fatal flaw since new
types are easy to define. However, since Java does not support operator
overloading, one cannot make such types behave like the primitive types
{\tt float} or {\tt double}. More important than syntactic convenience,
however, is that not having complex arithmetic in the language can
severely affect the performance of applications. This is because
compilers, as well as the JVM, will be unable to perform conventional
optimizations on complex arithmetic because they are unaware of the
semantics of the class. 

Since complex arithmetic is so pervasive it is necessary to establish community
consensus on a Java interface for complex classes \cite{VNI:1997:JLP}.

\subsection{Memory model}

Perhaps the biggest difference in developing numerical code in Java rather than
in Fortran or C results from Java's memory model.
Numerical software designers typically take information about the physical
layout of data in memory into account when designing algorithms to achieve high
performance.  For example, LINPACK \cite{Dongarra:1979:LUG} used
column-oriented algorithms and the Level 1 BLAS in order to localize memory
references for efficiency in paged virtual memory environments.  LAPACK
\cite{Anderson:1992:LUG} used block-oriented algorithms and the Level 2 and 3
BLAS to localize references for efficiency in modern cache-based systems.  The
ability to do this hinged on the fact that Fortran requires two-dimensional
arrays be stored contiguously by columns.

Unfortunately, there is no similar requirement for multidimensional arrays in
Java.  Here, a two-dimensional array is an array of one-dimensional arrays.
Although we might expect that elements of rows are stored contiguously, one
cannot depend upon the rows themselves being stored contiguously.  In fact,
there is no way to check whether rows have been stored contiguously after
they have been allocated.  The row-orientation of Java arrays means that, as in
C, row-oriented algorithms may be preferred over column-oriented algorithms.
The possible non-contiguity of rows implies that the effectiveness of
block-oriented algorithms may be highly dependent on the particular
implementation of the JVM as well as the current state of the memory manager.

The Java language has no facilities for explicitly referencing subvectors or
subarrays.  In particular, the approach commonly used in Fortran and C of
passing the address of an array element to a subprogram which then operates on
the appropriate subarray does not work.\footnote{In C one would explicitly pass
an address to the procedure, but address arithmetic does not exist in Java.  In
Fortran one passes an array element, which amounts to the same thing since all
parameters are passed by address, but scalars are passed by value in Java.}

\subsection{Java's vector class}

Despite its name, Java's vector class {\tt java.util.Vector} is not really
appropriate for numerics.  This class is similar in spirit to those found in
the Standard Template Library of C++, that is, they are merely containers which
represent objects logically stored in contiguous locations.

Because there is no operator overloading, access to vectors via this class must
be through a functional interface.  Also, {\tt Vector} stores
generic {\tt Objects}, not simple data types.  This allows a vector to contain
heterogeneous data elements --- an elegant feature, but it adds overhead, and,
unfortunately, complicates its use for simple data types.  To use a vector of
double, for example, one needs to use Java's wrapper {\tt Double} class and
perform explicit coercions.

Consider the difference between using native Java arrays,
\begin{center}
\begin{verbatim}
double x[] = new double[10];     // using native Java arrays
double a[] = new double[10];
...
a[i] =  (x[i+1] - x[i-1]) / 2.0;
\end{verbatim}
\end{center}
and Java's {\tt Vector} class,
\begin{center}
\begin{verbatim}
Vector x = new Vector(10);     // using Java's Vector class
Vector a = new Vector(10);
...
a.setElement(i, new Double((((Double) x.ElementAt(i+1)).doubleValue()
      - ((Double) x.ElementAt(i-1)).doubleValue()) / 2.0);
\end{verbatim}
\end{center}
Deriving a {\tt VectorDouble} class from {\tt Vector} which performed these
coercions automatically would clean the code up somewhat, but would introduce
even more overhead by  making each reference {\tt x[i]} a virtual functional
call.

\subsection{I/O facilities}

Java's JDK 1.1 defines over 40 I/O classes, many of them with only subtle
differences, making it difficult to choose the right one for a given task. For
example, the average Java user may not immediately recognize the difference
between parsing tokens from strings via {\tt StringTokenzier (String)} and {\tt
StreamTokenizer (StringReader(String))}.

Ironically, despite these numerous I/O classes there is little support for
reading floating point numbers in exponential notation. Even if one
resorts to low-level parsing routines to read floating point numbers, the
internal class {\tt java.io.StreamTokenizer} parses ``2.13e+6'' as four
separate tokens (``2.13'', ``e'', ``+'', ``6.0''), even when the {\tt
parseNumbers()} flag is set. Furthermore, no formatted output facilities
are provided, making it very difficult to produce readable tabulated
output. 

\subsection{Other inconveniences}

A number of other conveniences which library developers have come to depend
upon are not available in Java.  {\em Operator overloading}, which would be
particularly useful for user-defined matrix, vector, and array classes, as well
as for complex arithmetic, would be quite useful.  Finally, {\em templates},
such as those in C++, would eliminate the need to create duplicate functions
and classes for double, float, complex, etc.  The loss of compile-time
polymorphism can also lead to inefficiencies at run-time.  While these
omissions are not fatal, they significantly increase the burden of numerical
library developers.

Extensions to Java which provide such facilities are under development by
various groups, implemented as either precompilers producing pure Java, or as
compilers for the JVM\footnote{See
\verb+http://grunge.cs.tu-berlin.de/$\sim$tolk/vmlanguages.html+.} This may
provide a reasonable approach for a Java-centric development environment for
scientific computing.


\section{JAVA PERFORMANCE ON NUMERICAL KERNELS}

For numerical computation, performance is a critical concern.  Early experience
with Java interpreters has led to the common perception that applications
written in Java are slow.  One can get good performance in Java by using native
methods, i.e. calls to optimized code written in other languages
\cite{Bik:1997:NNL}.  However, this comes at the expense of portability, which
is a key feature of the language.  Advances in just-in-time (JIT) compilers,
and improvements in Java run-time systems have changed the landscape in recent
months, suggesting that Java itself may indeed be suitable for many types of
numerical computations.  In this section we provide a variety of benchmarks
which provide a brief glimpse of the current performance of Java for simple
numerical kernels.

\subsection{Run-time array bounds checking}

Run-time array bounds checking is a requirement of the JVM, and improved
reliability of applications makes this feature very desirable.  Fortunately,
this does not necessarily imply significant performance degradation in
practice.  Modern processors have the ability to overlap index checks with
other computation, allowing them to cost very little.  Experiments performed in
C++ using the NIST Template Numerical Toolkit
\cite{Pozo:1997:TNT} on a Pentium Pro with the Watcom C++ compiler (version
10.6) show that array bounds checking can add as little as 20\% overhead.

\subsection{Elementary kernels}

Timings for several BLAS 1 kernels with various unrolling strategies in several
environments are presented in Table \ref{Level1Kernels}.  The baseline kernel
{\tt daxpy} (unroll 1) is written with no optimizations, i.e.
{\small\begin{verbatim}
public static final void daxpy(int N, double a, double x[], double y[]) {
   for (int i=0; i<N ; i++) y[i] += a*x[i]; }
\end{verbatim}}
\noindent
The unroll 4 variant of daxpy uses the kernel
{\small\begin{verbatim}
     y[i  ] += a * x[i  ];  y[i+1] += a * x[i+1];
     y[i+2] += a * x[i+2];  y[i+3] += a * x[i+3];
\end{verbatim}}
\noindent
while the unroll 4-inc variant uses
{\small\begin{verbatim}
     y[i] += a * x[i]; i++;  y[i] += a * x[i]; i++;
     y[i] += a * x[i]; i++;  y[i] += a * x[i]; i++;
\end{verbatim}}
\noindent
The latter can provide performance improvements if the JVM's {\tt inc} opcode
is used.  The {\tt ddot} schemes are similar.

The data in Table \ref{Level1Kernels} show that Microsoft's SDK 2.0, for
example, appears to deliver about half the performance of C code for daxpy
and ddot on the Pentium II. This is encouraging. The results also indicate
that unrolling can favorably affect the performance of numerical kernels
in Java, but that the effect varies widely among JVMs. (No JIT is
currently available for the Sun JDK 1.1 under Linux.) 

\begin{table}
\caption{Performance of BLAS level 1 kernels in various environments. Vector
length is 200. Also varied is the depth of loop urollings.  Results in Mflops.
\label{Level1Kernels}}
\begin{center}
\begin{tabular}{|l|rrrr|rrr|}
\hline
  & \multicolumn{4}{c|}{\bf daxpy} & \multicolumn{3}{c|}{\bf ddot} \\
\cline{2-8}
\multicolumn{1}{|c|}{\bf Environment} & \multicolumn{4}{c|}{\bf Unroll depth} & \multicolumn{3}{c|}{\bf Unroll depth} \\
   & \multicolumn{1}{c}{1} 
   & \multicolumn{1}{c}{4} 
   & \multicolumn{1}{c}{4-inc} 
   & \multicolumn{1}{c|}{8} 
   & \multicolumn{1}{c}{1} 
   & \multicolumn{1}{c}{4} 
   & \multicolumn{1}{c|}{8} \\
\hline
{Pentium II, 266 MHz, Intel BLAS, Win95} &
96$^\dagger$ &  &  &  & 193$^\dagger$ &  &  \\
{Pentium II, 266 MHz, gcc 2.7.1 -O3, Linux} &
88.1 & 134.2 & 120.0 & 132.5 & 147.1 & 147.1 & 148.1 \\
{Pentium II, 266 MHz, Microsoft SDK 2.0, Win95} &
45.0 & 67.0 & 80.0 & 81.0 & 41.0 & 80.0 & 81.0 \\
\hline
{Pentium Pro, 200 MHz, Sun JDK 1.1.3, Linux} &
10.4 & 14.6 & 15.7 & 14.7 & 13.5 & 21.6 & 22.0 \\
\hline
{SGI R10000, 194 MHz, java 3.0.1, IRIX 6.2} &
12.0 & 15.0 & 16.7 & 16.1 & 14.4 & 22.0 & 24.4 \\
{SGI R10000, 194 MHz, f77 -O3, IRIX 6.2} &
128.7 & 128.9 & 129.3 & 133.3 & 188.4 & 188.7 & 186.3 \\
\hline
\multicolumn{8}{l}{$^\dagger$actual loop unrolling strategy unknown}
\end{tabular} 
\end{center}
\end{table}
\begin{table}
\caption{Performance of matrix multiplication in C and Java. 
266 MHz Pentium II using Gnu C 2.7.2.1 (Linux) and Microsoft Java SDK
2.0 (Windows 95).  C=AB, where A is LxN and B is NxM. L=M=100. Results in
Mflops.\label{TableMM}}
\begin{center}
\begin{tabular}{|c|rr|rr|}
\hline
& \multicolumn{4}{|c|}{\bf Environments} \\
&
\multicolumn{2}{c}{\bf Gnu C} & 
\multicolumn{2}{c|}{\bf Microsoft Java} \\
{\bf Loop order} & \multicolumn{1}{c}{N=100} & 
  \multicolumn{1}{c}{N=16}
& \multicolumn{1}{c}{N=100} & 
  \multicolumn{1}{c|}{N=16} \\
\hline
(i,j,k) &  82.2 &  90.6 & 20.4 & 29.1  \\
(k,i,j) &  60.4 &  49.0 & 11.4 & 13.9  \\
(i,k,j) &  74.1 &  60.0 &  7.6 &  9.4  \\
\hline
\end{tabular}
\end{center}
\end{table}
\begin{table}
\caption{Effect of loop optimizations on matrix multiplication in Java.
266 MHz Pentium II using Microsoft Java SDK 2.0 (Windows
95).  C=AB, where A is LxN and B is NxM. L=M=100. Results in
Mflops.\label{TableMMopt}}
\begin{center}
\begin{tabular}{|c|rr|rr|}
\hline
& \multicolumn{4}{|c|}{\bf Loop Optimizations} \\
&
\multicolumn{2}{c}{\bf 1D indexing} & 
\multicolumn{2}{c|}{\bf plus unrolling} \\
{\bf Loop order} & \multicolumn{1}{c}{N=100} & 
  \multicolumn{1}{c}{N=16}
& \multicolumn{1}{c}{N=100} & 
  \multicolumn{1}{c|}{N=16} \\
\hline
(i,j,k) & 30.4 & 36.4 & 38.3 & 49.2 \\
(k,i,j) & 18.2 & 20.8 & 22.0 & 26.0 \\
(i,k,j) & 10.4 & 11.2 & 15.8 & 18.5 \\
\hline
\end{tabular}
\end{center}
\end{table}
\begin{table}
\caption{Performance of sparse matrix-vector multiply in Java and C.
266 MHz Pentium II using Microsoft Java SDK 2.0 and Watcom C 10.6
(Windows 95). Results in Mflops.\label{TableSparse}}
\begin{center}
\begin{tabular}{|lrr|cc|}
\hline
& & & \multicolumn{2}{c|}{\bf Environments} \\
{\bf Matrix} & {\bf Order} & {\bf Entries} &
{\bf Microsoft Java} & {\bf Watcom C} \\
\hline
WEST0156 &    156 &     371 & 33.7 & 43.9 \\
SHERMAN3 &  5,505 &  20,033 & 14.0 & 21.4 \\
MCFE     &    765 &  24,382 & 17.0 & 23.2 \\
MEMPLUS  & 17,758 & 126,150 &  9.1 & 11.1 \\
\hline
\end{tabular}
\end{center}
\end{table}

\subsection{Dense matrix multiply}

We next consider a straightforward matrix multiply loop, i.e.
\begin{verbatim}
      for (int i=0; i<L; i++)
        for (int j=0; j<M; j++)
          for (int k=0; k<N; k++)
             C[i][j] += A[i][k] * B[k][j];
\end{verbatim}
By interchanging the three {\tt for} loops one can obtain six distinct matrix
multiplication algorithms.  We consider three which contain row operations in
the innermost loop, denoted as (i,j,k), (k,i,j) and (i,k,j) according to the
ordering of the loop indices.  The first is the loop displayed above;
it computs each element of C in turn using a dot product. The
second sweeps through C N times row-wise, accumulating one term of the inner
product in each of C's elements on each pass.  The third uses a row-wise
daxpy as the kernel.

In Table \ref{TableMM} we display the result in Mflops for these kernels on a
266 MHz Pentium II using both Java SDK 2.0 under Windows 95 and C compiled with
the Gnu C compiler Version 2.7.2.1 under Linux.  The C kernels were
coded exactly as the loop above, and compiled with the options {\tt -O3
-funroll-loops}.  We consider the case L=N=M=100, as well as the case where
L=M=100 and N=16.  The latter represents a typical rank K update in a
right-looking LU factorization algorithm.

In Table \ref{TableMMopt} we explore effect of two additional loop
optimizations for this kernel in Java on the Pentium II. In the first case
we attempt to reduce overhead using one-dimensional indexing. That is, we
assign rows to separate variables (e.g., {\tt Ci[j]} rather than {\tt
C[i][j]}), while in the second we use both one-dimensional indexing and
loop unrolling (to a level of 4). 

\begin{figure}
\centerline{\psfig{file=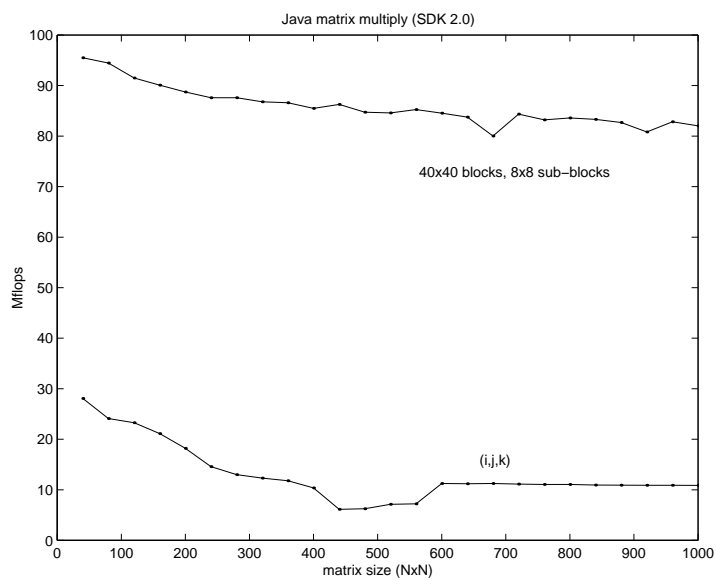,height=3in}}
\caption{\label{blockMMfig}Blocked versus Unblocked Matrix Multiplication 
in Java.}
\end{figure}

Efficient implementations of the Level 3 BLAS in Fortran use {\em blocked
algorithms} to make best use of cache.  Because the memory model used in Java
does not support contiguously stored two-dimensional arrays, there has been
some speculation that such algorithms would be less effective in Java.  We have
implemented a blocked variant of the (i,j,k) matrix multiply algorithm above
with two-level blocking: $40 \times 40$ blocks with $8 \times 8$ unrolled
sub-blocks. Figure \ref{blockMMfig} compares the simplistic (i,j,k) algorithm
with its blocked variant for matrices of size $40 \times 40$ to $1000 \times
1000$ on a 266 MHz Pentium II system running Microsoft Java SDK 2.0 under
Windows 95.  The performance of the blocked algorithm is far superior to the
unblocked algorithm, achieving 82.1 Mflops for the largest case.  This is only
10\% slower than the smallest case which fits completely in Level 1 cache.

The performance of the blocked algorithm is somewhat dependent on details
of the implementation.  We observed a subtle tradeoff, for example, between
including more temporary variables and the use of explicit indexing.  The
choice of blocksize also has a big effect; selecting $64 \times 64$ blocks
only yields 63 Mflop performance for the $1000 \times 1000$ case, for example.
This indicates that it may be necessary to have a standard method call which
returns the optimal blocksize for the current virtual machine.

These results indicate that Java performance is still inferior to that
obtained from C for Level 3 BLAS operations, but that optimized Java can
approach half the speed of C. (Note that the highly optimized Intel BLAS
run this kernel (dgemm) at about 190 Mflops for L=M=N=200.) Also, relative
performance of kernels in Java may be quite different than in C. As
expected, however, kernels based on dot products may be preferable in Java,
and strategies such as unrolling, one-dimensional indexing and blocking
will help, although the strength of the effect will undoubtably be highly
dependent on the particular JVM and JIT compiler.

\subsection{Sparse matrix-vector multiply}

Finally, we consider sparse matrix-vector multiplication based on a sparse
coordinate storage format \cite{Barrett:1994:TSL}. In particular, we
compare an implementation in Java with those based upon the NIST Sparse
BLAS Toolkit reference implementation \cite{Remington:1997:SBL} in C on a
266 MHz Pentium II running Windows 95. The test cases, taken from the
Harwell-Boeing collection \cite{Boisvert:1997:MMW,Duff:1989:SMT},
represent fairly small sparse matrices, but may provide an indication of
the relative performance of these languages on kernels which contain
indirect index computations. The results are presented in Table
\ref{TableSparse}. Note that the higher levels of performance for WEST0156
are due to the fact that the matrix is small enough to completely fit in
cache. The results indicate that optimized Java code which is rich in
indexing such as in sparse matrix operations can perform at from 65\% to
75\% of the speed of optimized C code. 


\section{NUMERICAL LIBRARIES IN JAVA}

Several development strategies exist for building numerical libraries in Java.
First, numerical classes can be coded directly in the language.  This is, of
course, labor-intensive, and could lead to less than optimal code due to
inefficiencies in the language.  Nevertheless, several groups have begun to
undertake such projects \cite{Russell:1997:PCP,Smith:1997:MML,VNI:1997:JLP}.  A
second option is to develop tools to translate existing Fortran
libraries into Java \cite{Casanova:1997:JAN,Fox:1997:PFJ}.  While this
provides easy access to a wealth of existing software, the severe mismatch
between Fortran and Java semantics is likely to lead to converted library
source which is unnatural and inefficient.  A third option is to use Java's
{\em native methods} facility to provide an interface to existing code in other
languages such as Fortran and C.  This requires one to develop a Java wrapper
to each library routine, although this is far simpler than recoding.  The
problem here, of course, is that Java's greatest advantage, its portability, is
compromised.

In this section we discuss issues involved in the design of numerical libraries
coded directly in the Java language.

\subsection{Basic Design Parameters}

A number of elementary design decisions must be made when
developing numerical libraries in Java.
\begin{itemize}
\item {\bf Precision.} 
      What floating-point precisions should be supported?
      Most numerical computations are currently carried out in IEEE
      double precision, and hence, support of double is necessary.
\item {\bf Naming.}
      What naming convention should be used for numerical classes?  
      Should long descriptive names or short less cumbersome names be used?
      Should {\tt Float} and {\tt Double} explicitly appear in class names
      to distinguish between precisions, as has been done in early numeric
      class libraries?
\item {\bf Vectors and Matrices.}
      Should native Java arrays be used instead of specialized classes for
      vectors and matrices?  Native Java arrays have the advantage of efficient
      processing and automatic array bounds checking.  If an elementary matrix
      class is devised, should matrices be represented internally as
      one-dimensional arrays to insure contiguity of data for block
      algorithms?  If this is done, how can we provide for efficient access to
      individual elements of arrays and vectors?  (Preliminary experiments with
      Microsoft SDK 1.1 using a five-point stencil kernel showed that
      use of explicit get and set methods in a matrix class was five times
      slower than using native Java arrays.)  Should indexing of vectors and
      matrices be 0-based or 1-based?  Should packed storage schemes be
      supported?  One can argue that storage is now so plentiful that for many
      problems the complexity of packed storage schemes for triangular and
      symmetric matrices is unnecessary in many cases.
\item {\bf Serializable classes.}
      Java provides a convention that allows for I/O of arbitrary objects.
      Classes which implement the {\tt Serializable} interface promise to provide
      standard utilities for input and output of their instances.  Should all
      numeric classes be made {\tt Serializable}?
\item {\bf Functionality.}
      How much functionality should be built into classes?
      Experience has shown that extensive object-oriented design frameworks 
      tend to restrict usability.  
\end{itemize}

Because the design of object-oriented frameworks for numerical computing is a
very difficult undertaking, and elaborate designs may, in fact, limit
usability by being too complex and specialized for many users, we propose that
a {\em toolkit} approach be taken to the development of numerical libraries in
Java.  A toolkit is a collection of ``raw'' classes which are unencumbered by
most of the trappings of object-oriented computing frameworks.  They provide a
rich source of numerical algorithms implemented mostly as static methods which
need not be explicitly instantiated to be used.  For simplicity and efficiency,
native Java arrays are used to represent vectors and matrices.  A toolkit
provides a low-level interface to numerical algorithms similar to what one
finds in C and Fortran.  Toolkits provide a source of basic numerical kernels
and computational algorithms which can be used in the construction of more
facile object-oriented frameworks for particular applications
\cite{Dwyer:1997:FPA}.

\subsection{Interface to the BLAS}

Because the Java memory model for multidimensional arrays is different than the
Fortran model, some consideration must be given to the meaning and
interpretation of function interfaces for matrix/vector kernels like the BLAS
which play a key role in any toolkit for numerical linear algebra.

As in C and C++, Java stores matrices by rows, without any guarantee that 
consecutive rows are actually contiguous in memory.  Java goes one step
further, however. Because there are no capabilities to manipulate pointers
directly, one cannot ``alias'' subvectors, or reshape vectors into matrices
in a direct and efficient manner.

For example, in Fortran, if the function {\tt SUM(N, X)} sums N elements from a
given vector X, then calling it as {\tt SUM(K, X(I))} sums the elements $x_i,
x_i+1, ..., x_i+K$. Unfortunately, no analogue exists in Java. We must
reference subvectors explicitly by describing the whole vector and its offset
separately, i.e. {\tt SUM(N, X, I)}.  

If we are to provide the same level of functionality as the Fortran and C
BLAS then we must provide {\em several} versions of each vector operation. For
example, the functionality of a Fortran BLAS with calling sequence
\begin{center}\tt
(..., N, X, INCX, ... )
\end{center}
would have to be supported by
\begin{itemize}
\item[] {\tt (..., int n, double x[], int xoffset, ... )}, and
\item[] {\tt (..., int n, double A[][], int Arow, int Acol, ... )},
\end{itemize}
\noindent
the former for a subvector, the latter for part of a column in a
two-dimensional array.
Thus, a Fortran call to manipulate a subvector such as
\begin{center}\tt
CALL DAXPY(N, ALPHA, X(I), 1, Y(I), 1)
\end{center}
would be realized in Java as
\begin{center}\tt
BLAS.daxpy(N, alpha, x, i, y, i)
\end{center}
whereas a Fortran call to manipulate part of the column of an array such as
\begin{center}\tt
CALL DAXPY(N, ALPHA, A(I,J), LDA, B(I,J), LDB)
\end{center}
would be realized in Java as
\begin{center}\tt
BLAS.daxpy(N, alpha, A, i, j, B, i, j)
\end{center}
One might also want to provide simplified, and more efficient, versions which
operated on entire vectors or columns, e.g.,
\begin{itemize}
\item[] {\tt (..., double x[], ...)} and
\item[] {\tt (..., double A[][], int Acol, ... )},
\end{itemize}
the former for a vector, the latter for a column in a two-dimensional array.

Similarly, Level 2 and Level 3 BLAS which refer to matrices in Fortran as
\begin{center}\tt
(..., N, M, A, LDA, ...)
\end{center}
would require explicit offsets in Java to support operations on subarrays as in
\begin{center}\tt
(..., int n, int m, double A[][], int Arow, int Acol, ... )
\end{center}
whereas routines which manipulate whole Java arrays need only have
\begin{center}\tt
(..., double A[][], ... )
\end{center}

It is clear that providing efficient and capable linear algebra kernels in
Java requires much more coding than in Fortran or C.

\subsection{Interfaces}
\label{sectionInterfaces}

It is also necessary to identify common mathematical operations to be defined
as Java interfaces.  An interface is a promise to provide a particular set of
methods.  User-defined objects implementing a well-defined interface can then
be operated on by library routines in a standard way.

Interfaces to generic mathematical functions are needed, for example, in order
to be able to pass user-defined functions to zero finders, minimizers,
quadrature routines, plotting routines, etc.  If the following existed,
\begin{verbatim}
     public interface UnivariateFunction {
        double eval(double x);}
\end{verbatim}
then instances of user-defined classes implementing UnivariateFunction could be
passed as arguments to zero finders, which in turn would use the eval method to
sample the function.  Many variants of mathematical function interfaces would be
required.  For example, it would also be necessary to define interfaces for
bivariate, trivariate and multivariate functions.  It would also be necessary
to define interfaces for transformations from $R^m$ to $R^n$.  Versions for
complex variables would also be required.

Interfaces are necessary to support iterative methods for the solution of
sparse linear systems.  These would define standard method invocations for
operations such as the application of a linear operator (matrix-vector
multiply) and preconditioner application, thus allowing the development of
iterative solvers that are independent of matrix representation.


\section{THE JAVA NUMERICAL TOOLKIT} 

In order to promote widespread reuse, community defined standard class
libraries and interfaces are needed for basic mathematical operations.  
To promote the development of such class libraries, we have begun the
construction of the Java Numerical Toolkit (JNT).  JNT will contain
basic numerical functions and kernels which can be used to build more
capable class libraries.  In particular, the initial version of
JNT includes
\begin{outline}
\item elementary matrix/vector operations (BLAS)
\item dense LU and QR matrix factorizations
\item dense linear systems and least squares problems
\item sparse linear systems using iterative methods
\item elementary and special functions, such as sign and Bessel functions
      $I_0$, $I_1$, $J_0$, $J_1$, $K_0$, $K_1$, $Y_0$, $Y_1$
\item random number generators
\item solution of nonlinear equations of a single variable
\end{outline}
The toolkit includes interface definitions to support mathematical
functions and linear solvers as in Section \ref{sectionInterfaces}.
JNT will also include
\begin{outline}
\item solution of banded linear systems
\item dense eigenvalue problems
\item support for a variety of specialized matrix formats
\item additional special functions, such as hyperbolic functions,
      the error function, gamma function, etc.
\item one-dimensional quadrature rules
\end{outline}

The toolkit has been initially developed using the floating-point type {\tt
double}.  Class and method names will not include the word Double (i.e., double
will be the toolkit default).  Versions based upon {\tt float} will be defined
and developed later if there is sufficient demand.

We are using the initial version of the toolkit to develop
prototype user-level classes for numerical linear algebra.  These
will include explicit dense matrix classes which will contain class
variables for matrix factorizations.   When a user solves a linear
system with such a matrix object, a factorization would automatically
be computed if necessary and would be available for future use,
without explicit action or knowledge of the programmer.
 
The Web page for the JNT project is {\tt http://math.nist.gov/jnt/}.


\section{CONCLUSIONS} 

The high level of portability, along with support for GUIs and network-based
computing provided by Java is attracting interest from the scientific computing
community.  The Java language itself provides many facilities needed for
numerical computing, but many others are lacking, such as complex arithmetic,
operator overloading, a clear memory model, and formatted I/O.  These will lead
to much additional effort on the part of programmers, and brings the ability
to achieve high levels of performance in some areas into doubt.  On the other
hand, rapid progress is being made in the development of JIT compilers, and the
performance level of many Java systems are improving (delivering as much as
25-50\% of optimized Fortran and C for key kernels in some cases).  A major
impediment to quick progress in this area is the lack of basic mathematical
software which is plentiful in other environments.  The construction of
basic numerical toolkits for Java needs to be undertaken to bootstrap the
development of more sophisticated numerical applications and to provide a basis
for the development of community supported standard numerical class libraries
and interfaces.


\section*{ACKNOWLEDGMENTS} 

This paper is a contribution of the National Institute of Standards and
Technology and is not subject to copyright.
Certain commercial products are identified in this paper in order to
adequately document computational experiments.
Identification of such products does not constitute endorsement by NIST,
nor does it imply that these are the most suitable products for the task.

This work is supported in part by Defense Advanced Research
Projects Agency under contract DAAH04-95-1-0595, administered by the U.S.~Army
Research Office.


\end{document}